\documentclass[12pt, letterpaper]{article}

\usepackage{amsmath}
\usepackage{amsfonts}
\usepackage{amssymb}
\usepackage{graphicx}

\usepackage{color}

\usepackage{amsmath, amssymb, graphics, epsfig, graphicx}
\usepackage{epsf}
\usepackage{epstopdf}
\usepackage {amssymb}
\newcommand{\nc}{\newcommand}
\nc{\ba}{\begin{eqnarray}}
\nc{\ea}{\end{eqnarray}}
\newcommand\be{\begin{equation}}
\newcommand\ee{\end{equation}}

\newcommand{\calR}{{\cal{R}}}
\newcommand{\calH}{{\cal{H}}}

\newcommand{\bea}{\begin{eqnarray}}
\newcommand{\eea}{\end{eqnarray}}

\newcommand{\bfx}{{\bf{x}}}

\begin{document}

\vspace{5mm}
\vspace{0.5cm}
\begin{center}

\def\thefootnote{\fnsymbol{footnote}}

{\Large  Cosmological Perturbations and the Weinberg Theorem}
\\[0.5cm]

{ Mohammad Akhshik$^{1, 2}$, Hassan Firouzjahi$^{1}$,  Sadra Jazayeri$^{1}$}

{\small \textit{$^1$School of Astronomy, Institute for Research in Fundamental Sciences (IPM) \\ P.~O.~Box 19395-5531, Tehran, Iran
}}\\

{\small \textit{$^2$Department of Physics, Sharif University of Technology, Tehran, Iran
}}\\
\vspace{0.1cm}
{\small {e-mails:  m.akhshik@ipm.ir, ~
 firouz@ipm.ir, ~ sadraj@ipm.ir
}}\\

\end{center}

\vspace{.8cm}

\hrule \vspace{0.3cm}


\begin{abstract}

The celebrated Weinberg theorem in cosmological perturbation theory states that there always exist two adiabatic scalar modes in which the comoving curvature perturbation is conserved on super-horizon scales. In particular, 
when the perturbations are generated from a single source, such as in single field models of inflation,  both of the two allowed independent solutions are adiabatic and conserved on super-horizon scales. There are few known examples in literature 
which violate this theorem. We revisit the theorem and specify the loopholes in some technical assumptions  which violate  the  theorem in models of non-attractor inflation, fluid inflation, solid inflation and in
the model of pseudo conformal universe. 

\end{abstract}
\vspace{0.5cm} \hrule
\def\thefootnote{\arabic{footnote}}
\setcounter{footnote}{0}
\newpage
\section{Introduction}

Cosmological perturbations theory is the vital tool to connect the predictions of perturbations generated from seed 
quantum fluctuations in early universe, such as during inflation, to late time cosmological observations such as
cosmic microwave background (CMB) or large scale structures (LSS).  After inflation ends, the universe enters into the violent phase of reheating and the follow up radiation and matter dominated eras with different sources of energy and matter constituents. However, the fact that there exists adiabatic perturbations which are conserved on super-horizon scales is  a powerful tool to  connect the large scale fluctuations in CMB or LSS to the corresponding curvature perturbations generated during inflation  when the mode of interest leaves the  horizon. 

It is well-known that the comoving curvature perturbation $\calR$ or the curvature perturbations on surface of constant energy density $\zeta$ are conserved on super-horizon scales in models of single field slow-roll inflation, for a review see
\cite{Weinberg:2008zzc, Mukhanov:2005sc,  Lyth:2009zz, Bassett:2005xm}. Weinberg has generalized this conclusion to a broad class of cosmological perturbations in early universe \cite{Weinberg:2008zzc, Weinberg:2003sw}.  The celebrated Weinberg theorem states that whatever the content of the universe, the comoving curvature perturbations in Newtonian gauge always has two adiabatic modes which are frozen on super-horizon scales, corresponding to $k/a \ll H$ in which $a$ is the cosmic scale factor, $H$ is the Hubble expansion rate and $k$ is the comoving wave-number (in Fourier space). This theorem also states that in addition 
there is one tensor mode which is  conserved on super-horizon scales. In our studies here, we shall concentrate on scalar perturbations.

In particular,  Weinberg's theorem has strong implications for models in which 
perturbations are generated from a single source, such as in models of single field  inflation. In these models, the counting of independent degrees of freedom indicate that we have only two independent modes of curvature perturbations. Consequently,  Weinberg's theorem imply that both of these two modes should be conserved  on super-horizon scales  in  single field models. The theorem states that  the dominant mode is the usual conserved mode in single field inflation models while the other adiabatic mode is actually $\calR_k=0$. Of course, these conclusions conform with the know results in single field slow roll inflation models  as mentioned above (more precisely, in single field slow roll models the decaying mode approaches $\calR_k=0$). 
However, there are known examples in literature such as models of  non-attractor inflation, fluid inflation, solid inflation, pseudo conformal universe  and  Galilean Genesis in which the curvature perturbation is not frozen on super-horizon scales. For example it is known that in models of non-attractor  inflation the usual would-be decaying mode is actually the growing mode and 
$\calR$ grows like $a^{3}$  \cite{Namjoo:2012aa, Chen:2013aj, Chen:2013eea, Akhshik:2015nfa, Mooij:2015yka}.  Logically, therefore,  one is led to ask how these models evade  Weinberg's theorem. The goal of this work is to shed some light on this question. We revisit the mechanism in which this theorem is proved and specify the loopholes in some technical assumptions required in the theorem which are violated in these scenarios. There are some generic features on the violation of these technical assumptions which are shared in these models but we shall study each model independently to specify the exact nature of the violation of the theorem.

\section{A brief review of Weinberg's theorem}
\label{Weinberg-section}

In this section we briefly review the Weinberg's theorem which is independent of model (i.e. without assuming  scalar fields etc.). For a more extensive review see \cite{Weinberg:2008zzc, Weinberg:2003sw}.

We are interested in scalar perturbations of the metric and matter sources. The scalar sector of metric perturbations 
in the Newtonian gauge has the  following form
\ba
ds^2 = -(1+ 2 \Phi( t , \bfx) ) dt^2  + a(t)^2 \left( 1 - 2\Psi(t, \bfx)\,     \right) d \bfx^2 \, ,
\ea
in which $\Phi$ and $\Psi$ are the Bardeen potentials. The advantage in using the Newtonian gauge in the analysis 
of  \cite{Weinberg:2008zzc, Weinberg:2003sw} is that  this gauge leaves no residual gauge symmetry except for the mode with the zero wavenumber, $k=0$. This was crucially used in the proof of the theorem. 

Let us start with the homogeneous FRW background and then consider the solutions of the perturbed Einstein fields equations which are homogeneous but time-dependent: $\Phi= \Phi(t)$ and $\Psi=\Psi(t)$. Of course, they are not physical solutions by themselves as in general they may be removed by a  coordinate transformation, $x^\mu \rightarrow x^\mu+ \epsilon^\mu(t, \bfx)$.  The goal is to see under what conditions a subset of  these solutions can be extended to  non-zero wavenumber which satisfy all Einstein's equations. If so, then these subset of solutions represent physical solutions.    

As demonstrated in 
\cite{Weinberg:2008zzc, Weinberg:2003sw} one concludes that there is always a spatially homogenous solution 
to the set of perturbed Einstein equations in Newtonian gauge in which 
\ba
\label{hom1}
\Psi(t) &=& H \epsilon(t) - \frac{\omega_{ii}}{3} \quad , \quad \Phi(t)= -\dot \epsilon(t)  \\
\label{hom2}
\delta p&=& -\dot p \, \epsilon(t) \quad , \quad \delta \rho = -\dot \rho \,  \epsilon(t) , \quad ,\quad
\delta u = \epsilon(t) \quad, \quad \pi^S =0  \, ,
\ea
where $\delta \rho$ and $ \delta p$ represent respectively the perturbed  energy density and pressure, $\delta u
$ is the perturbed velocity potential and $\pi^S$ is the anisotropic inertia (pressure) term. In addition, $\epsilon(t)$
is a function encoding the time-dependent part of $\epsilon_0(t, \bfx)$ in the coordinate transformation 
$x^\mu \rightarrow x^\mu+ \epsilon^\mu(t, \bfx)$ and $\omega_{ij}$ is a constant matrix (note that $\omega_{ii}$
is the trace of $\omega_{ij}$).

As mentioned above, the solution given in Eqs. (\ref{hom1}) and (\ref{hom2}) are not physical in general. They become physical if they can be promoted to non-zero wave-numbers. In other words, the solutions in   Eqs. (\ref{hom1}) and (\ref{hom2}) become physical if they also satisfy the Einstein fields equations when $k \neq 0$. 
Imposing that    Eqs. (\ref{hom1}) and (\ref{hom2})  also satisfy the inhomogeneous perturbed Einstein equations 
one obtain two sets of independent physical  solutions. The first set of solution is given by
\ba
\label{set1}
\Psi&=& \Phi = \calR \left[ -1+ \frac{H(t)}{a(t)} \int_\tau^t a(t') d t'  \right]  \nonumber\\
\frac{\delta p}{\dot p}&=&  \frac{\delta \rho}{\dot \rho} =- \delta u = -\frac{\calR}{a(t)}  \int_\tau^t a(t') d t' \, ,
\ea
in which $\calR$ is the comoving curvature perturbation which is also conserved, $\calR = \omega_{ii}/3$. 

The second class of the physical solution is obtained to be
\ba
\label{set2}
\Psi &=& \Phi = \frac{C H(t)}{a(t)} \nonumber\\
\frac{\delta p}{\dot p}&=&  \frac{\delta \rho}{\dot \rho} =- \delta u = -\frac{C}{a(t)} \, ,
\ea
in which $C$ is a constant.  Furthermore, for this mode $\calR =0$. 

We note that in both classes of solutions all scalar quantity $s$ such as
$\rho$ or $p $ have equal value for $\delta s/\dot s$, i.e. $\delta \rho/\dot \rho= \delta p/\dot p$. For this reason these solutions are called adiabatic. In addition, in order to simplify our presentation of the theorem, 
we implicitly assumed that there is no anisotropic stress, $\pi^S=0$ and consequently $\Phi =\Psi$. However, as in \cite{Weinberg:2003sw}, one can
extend these analysis to more general case in which $\pi^S \neq 0$.

This summarizes the statement of the theorem. The  details of assumptions and the derivations employed 
 in \cite{Weinberg:2008zzc, Weinberg:2003sw} seem to leave no loophole. However, there are two technical assumptions which may not be justified in general. The first technical assumption is that the set of perturbed Einstein equations are regular at $k=0$ so the transition from the gauge mode $k=0$ to the physical mode with
 $k\neq 0$ but with $k\rightarrow 0$ can be made continuously. The necessity of this technical assumption was already  mentioned in \cite{Weinberg:2003sw} (see also \cite{Berezhiani:2014kga}) and the fact that this technical assumptions may be invalidated in some certain cases. As we shall see a particular example in which this technical assumption is violated is 
the model of solid inflation. 

However, a more subtle and somewhat hidden point in the proof of \cite{Weinberg:2008zzc, Weinberg:2003sw}  is the extent to which one can take the limit  $k\rightarrow 0$ arbitrarily for the super-horizon mode, without causing difficulties. The super-horizon  condition is $k/a H \ll 1$. So whenever we take  $k\rightarrow 0$ when dealing with the  Einstein equations  we actually mean  the extent to which $k/a H $ goes to zero. Now suppose the fields equations or the constraints are written such that 
 \ba
 \label{example}
 \alpha(t) y_1 + \frac{k^2}{a ^2 H^2} y_2 = \beta(t) y_3 \, ,
 \ea
in which $\alpha(t)$ and $\beta(t)$ are functions of  background quantities such as $H$, $\dot H$ etc but independent of $k$,  and $y_i$ collectively represents some physical fields. As we shall see, the Poisson equation is a constraint   like the above equation, see Eq. (\ref{Poisson2}). Now when we take $k\rightarrow 0$ as the definition of super-horizon limit we actually mean  
$\frac{k^2}{a ^2 H^2} \rightarrow 0$ to arbitrary extent. However, this should be compared with the coefficients 
$\alpha(t)$ or $\beta (t)$. For example, if the coefficient $\alpha(t)$ approaches zero more faster than $1/a^2$, then taking $k\rightarrow 0$ as the criteria to turn on a physical super-horizon mode from a  pure gauge mode $k=0$ is ill-defined. As we shall see this is exactly what happens in models of non-attractor inflation  in
which $\alpha(t)$ falls off like $a^{-6}$, much  faster then the combination $\frac{k^2}{a ^2 H^2}$. In this situations the proof of \cite{Weinberg:2008zzc, Weinberg:2003sw} is not expected to go through and  the results of \cite{Weinberg:2008zzc, Weinberg:2003sw} are violated in one way or another.

\section{Non-attractor inflation}
\label{non-attractor-sec}
In this section we study in details how Weinberg's theorem is violated in models of non-attractor inflation.

Let us first briefly review the models of non-attractor inflation. These models are proposed as a counter example which violate the Maldacena's single field non-Gaussianity consistency condition \cite{Maldacena:2002vr, Creminelli:2004yq}. In its simplest realization \cite{Namjoo:2012aa} (see also \cite{Kinney:2005vj}), the model consists of a scalar field $\phi$ rolling in a flat potential $V=V_0$. From the background field equations we obtain
$\dot \phi \propto a^{-3}$ while the first slow-roll parameter $\epsilon \equiv -\dot H/H^2$ falls off like $a^{-6}$.
As studies in \cite{Namjoo:2012aa} a scale-invariant curvature perturbation with $n_s=1$ can be obtained with the
second slow-roll parameter $\eta \equiv \dot \epsilon/\epsilon H \simeq -6$. The large deviation of $\eta$ from the usual slow-roll condition is a manifestation of the fact that the potential is exactly flat and $\epsilon$ falls off exponentially 
during inflation. The crucial effect in non-attractor model is that the dominant curvature perturbation is not frozen
on super-horizon scales and $\calR$ grows like $a^3$. However, note that we still have the constant mode solution 
for $\calR$ which is now the sub-leading mode. Putting it another way, the would-be decaying mode in conventional slow-roll inflation is now actually the growing mode while the would-be dominant  mode  in slow-roll models, corresponding to  $\calR =\mathrm{constant}$, is here the sub-leading mode. Now comparing with  Weinberg's theorem, we recover the mode $\calR =\mathrm{constant}$. However, we do not recover the other solution $\calR =0$ and instead we get $\calR \propto a^{3}$. This obviously calls for an inspection as how the Weinberg theorem is violated in this setup. 

There are two important comments in order. The first comment is that  the fact that $\calR$ is not frozen on super-horizon scales is the key to violate the single field non-Gaussianity consistency condition. Indeed, if $\calR$ was frozen on super-horizon scales then by a change of coordinate $x^i \rightarrow e^{\calR} x^i$ one could eliminate $\calR$ completely yielding a zero value for the non-Gaussianity parameter $f_{NL}$ in the squeezed limit. The second comment is that the model, as proposed, suffers from the graceful exit problem as there is no mechanism to terminate inflation. However, in a more realistic situation one can imagine that towards the end
of inflation a mechanism like waterfall phase transition happens terminating inflation efficiently. This can be achieved by a heavy
waterfall field which has no contribution in curvature perturbation as in models of hybrid inflation. 

The above simple  non-attractor model was extended to more interesting case in the context of K-inflation 
in which the potential is not flat and the scalar perturbations have a non-trivial sound speed $c_s$  \cite{ Chen:2013aj, Chen:2013eea}, see also \cite{Motohashi:2014ppa}. The non-Gaussianity parameter $f_{NL}$ in the squeezed limit is given by $f_{NL}= 5(1+ c_s^2)/4 c_s^2$ which clearly violates Maldacena's consistency condition. 

For the later reference, it is helpful to calculate the relation between $\Phi$ and $\calR$ given in Eq. (\ref{set1}) for the first adiabatic mode in Weinberg's theorem in non-attractor model. 
With an integration by parts the relation between $\Phi$ and $\calR$ is obtained to be
\ba
\Phi = \calR \left[ -1 + \frac{H}{a} \left( \frac{a}{H} + \int \frac{a d H}{H^2} \right)
\right]  \, .
\ea
Now taking $\epsilon =-\dot H/H^2 \propto \tau^6 $, and to leading order in $\epsilon $,  $H \tau \simeq -1$, the above
integral can be cast into an integral over $\tau $ in the form of $\int d \tau \tau^4$ yielding
\ba
\label{R-Phi-ratio}
\Phi= \frac{\epsilon}{5} \calR \,   .
\ea
We emphasis again that the above relation between $\Phi$ and $\calR$ is valid only for the first mode in Weinberg's theorem  given in Eq. (\ref{set1}) which will be used in subsequent analysis. 

Below we demonstrate the violation of the theorem in simple model of non-attractor inflation \cite{Namjoo:2012aa} with $V(\phi)=V_0$ in three different methods. 
In the first method, we obtain the second order differential equation for $\calR$ and specify how the theorem is violated. In the second method, we solve the sets of Einstein equations to obtain $\Phi$ directly and look at its
super-horizon limit $k/a H \ll 1$ or alternatively $k \tau \rightarrow 0$ in which $\tau$ is the conformal time 
related to physical time via $d\tau = dt/a(t)$. In the third method, we construct the solution first  in the comoving 
gauge and then calculate $\Phi$ in Newtonian gauge which enables us to view the violation of the theorem from a different perspective.  

\subsection{An equation for $\calR$}
\label{R-method}

We work in the Newtonian gauge and set $\Psi =\Phi$ as there is no anisotropic inertia. 
Going to Fourier space,  the set of perturbed Einstein equation to be solved  are
\ba
\label{0i}
\dot \Phi + H \Phi =4 \pi G  \dot \phi \, \delta \phi \\
\label{KG}
 \delta  \ddot\phi + 3 H \delta   \dot\phi  + \frac{k^2}{a^2} \delta \phi = 4 \dot \phi \dot \Phi \, ,
\ea
supplemented with the constraint equation (the Poisson equation)
\ba
\label{Poisson}
\left(\dot  H +  \frac{k^2}{a^2} \right) \Phi =  4 \pi G \left(  -\dot \phi \delta \dot \phi + \ddot \phi \delta \phi
\right) \, ,
\ea
in which a dot indicates the derivative with respect to cosmic time $t$ and $G$ is the Newton constant.

It is more convenient to work with the velocity potential $\delta u= - \delta \phi/\dot \phi$ in which Eqs. (\ref{0i})
and (\ref{Poisson}) are cast into
\ba
\label{0i-2}
\dot \Phi + H \Phi = - \epsilon H^2 \delta u \, ,
\ea
and
\ba
\label{Poisson2}
\left(  \epsilon - \frac{k^2}{a^2 H^2} \right) \Phi = -  \epsilon \delta \dot  u \, 
\ea
in which $\epsilon =-\dot H/H^2 = 4\pi G \dot \phi^2/H^2$. 

As promised before, Eq. (\ref{Poisson2}) has the form of Eq. (\ref{example}) and we can guess how the theorem 
in \cite{Weinberg:2008zzc, Weinberg:2003sw} may be violated.  If we take the {\it arbitrary} 
mathematical limit $k\rightarrow 0$ then the second term in Eq. (\ref{Poisson2}) can be discarded and we obtain the relation $\Phi = -\delta \dot u$ which is the starting point in   \cite{ Weinberg:2003sw} when proving the theorem for the scalar fields. 
In usual situations, such as in slow-roll models, in which $\epsilon$ is nearly constant,  taking the super-horizon limit simply as $k \rightarrow 0$ is safe justifying neglecting the second term in Eq. (\ref{Poisson2}). However, in the non-attractor model we have $\epsilon \propto a^{-6}$ so the first term in Eq. (\ref{Poisson2}) falls off much faster than the second term. On the other hand, when we take $k\rightarrow 0$ we actually rely on the fact that
$a(t)$ expands exponentially so $k/ a H$ falls off  quickly for a {\it given} $k$. This was the trick to turn on the physical solution from the pure gauge mode $k=0$  in \cite{Weinberg:2008zzc, Weinberg:2003sw}. Now in the non-attractor models, with the first term in Eq. (\ref{Poisson2}) falling much faster than the term containing $k^2$, then taking $k \rightarrow 0$ as the criteria for super-horizon mode is ill-defined.  Surprisingly, the would be decaying term in Eq. (\ref{Poisson2}) (the term containing $k^2$)  now is the leading term. For this reason, we keep both terms in bracket in Eq. (\ref{Poisson2}) without dropping the term 
containing  $k^2$. 

The comoving curvature perturbation $\calR$  is given by 
\ba
\label{R-def}
\calR = H \delta u - \Phi \, .
\ea
Plugging this into the conservation equation (\ref{0i-2}) yields
\ba
\label{R-u}
H \delta \dot u + H^2 \delta u = \dot \calR + H \calR \, .
\ea
Now we manipulate Eqs. (\ref{Poisson2}), (\ref{R-u}) to obtain
\ba
\label{R-u2}
\delta u = \frac{\calR}{H} + \frac{a^2 \epsilon}{k^2} \dot \calR 
\ea
and
\ba
\label{Phi-R}
\Phi=  \left( \frac{\epsilon a^2 H^2}{k^2} \right)  \frac{\dot \calR}{H} \, .
\ea
The above equations show a non-trivial interplay between $\dot \calR$ and $k^{-2}$. Indeed, taking the mathematical limit $k^2=0$ requires that $\dot \calR =0$ for the  equations to be consistent. This brings us to
the conclusion of  \cite{ Weinberg:2003sw}.  

Now, with $\delta u$ and $\Phi$ expressed in terms of $\calR$ and $\dot \calR$ in Eqs. (\ref{R-u2})
and (\ref{Phi-R}), we can cast the remaining equation (\ref{KG}) into a second order differential equation
for $\calR$. With some long but otherwise simple manipulations we obtain
\ba
\label{R-eq}
\partial_t \left(  a^3 \epsilon \dot \calR \right) +  k^2  \epsilon a \, \calR =0 \, .
\ea
This is a known equation for $\calR$ which can easily be obtained in other gauges, such as comoving gauge as
employed in  \cite{Namjoo:2012aa}. However, we went into long procedure of deriving Eq. (\ref{R-eq}) in Newtonian gauge in order to be on the same platform as in  \cite{Weinberg:2008zzc, Weinberg:2003sw} and in order to pin down the loophole in the technical assumption employed in  \cite{Weinberg:2008zzc, Weinberg:2003sw}  to prove the theorem. 

Now, the super-horizon limit in Eq. (\ref{R-eq}) can be taken without any problem.  The mathematical limit of taking $k \rightarrow 0$ as employed in  \cite{Weinberg:2008zzc, Weinberg:2003sw}  makes sense only in Eq. (\ref{R-eq}) in which the coefficient of $\dot \calR$, $a^3 \epsilon$, does not vanish faster than the coefficient of $k^2$.  This is opposite to the situation  in Eq. (\ref{Poisson2}) in which the first term in Eq. (\ref{Poisson2}) falls off much faster than the
second term  containing $k^2$.

Taking the super-horizon limit  of  Eq. (\ref{R-eq})  we obtain 
\ba
\label{R-sol}
\calR = C_1 + C_2 \int \frac{dt }{a^3 \epsilon} \, ,
\ea
in which $C_1$ and $C_2$ are two constants of integrations representing the two independent modes.
The mode represented by $C_1$ is the usual mode which  also exists in  \cite{Weinberg:2008zzc, Weinberg:2003sw}. The difference now is in the mode represented by $C_2$. In conventional slow-roll model in which $\epsilon$ is constant, this mode  decays and one approaches the other solution in \cite{Weinberg:2008zzc, Weinberg:2003sw} labeled  by $\calR =0$. However, in non-attractor model in which $\epsilon \propto a^{-6}$, this solution is the growing mode yielding $\calR \sim a(t)^3$ as observed in  \cite{Namjoo:2012aa}. 

\subsection{ The equation for $\Phi$}
\label{Phi-method}

Here we solve the Einstein equations in Newtonian gauge directly to obtain $\Phi$. The corresponding  equations involving the (00) and (ii) components of Einstein's equations, with $\Psi =\Phi$, are 
\ba
\label{Eins}
\ddot{\Phi}+7H\dot{\Phi}+(6H^2+2 \dot{H})\Phi +\frac{k ^2}{a^2}\Phi =-4 \pi G (\delta \rho-\delta P)
\\ 
3\ddot{\Phi}+9H\dot{\Psi}+6(H^2+\dot{H})\Phi -\frac{k^2}{a^2}\Phi=   4 \pi G (\delta \rho+3\delta P) 
\ea
while the (0i) equation is as given in Eq. (\ref{0i}). 

The general forms of $\delta \rho$ and $\delta P$ are given by 
\ba
\delta P=\dot{\phi}\delta \dot{\phi}-\dot{\phi}^2\Phi-V_{\phi}\delta \phi\\ 
\delta \rho=\dot{\phi}\delta \dot{\phi}-\dot{\phi}^2\Phi+V_{\phi}\delta \phi  \, .
\ea
Note the curious effect that in our simple non-attractor model with a constant potential, $V=V_0$, we obtain
$\delta \rho=\delta P= \dot{\phi}\delta \dot{\phi}-\dot{\phi}^2\Phi$. With $\delta \rho=\delta P$, Eq. (\ref{Eins})
can be solved directly without the need to 
solve for $\delta u$, $\delta \rho$ and $\delta P $ from other equations.  

Our goal is to find the solution of $\Phi$ from Eq. (\ref{Eins}) and then use this value of $\Phi$ to calculate $\calR$. Note that from Eq. (\ref{R-def}), and after eliminating 
$\delta u = -\delta \phi/\dot \phi $ using Eq. (\ref{0i}),  the relation between $\Phi$ and $\calR$ is 
\ba
\label{R-Phi2}
\calR = -\Phi + \frac{H}{\dot H} \left( \dot \Phi + H \Phi \right) =  -\Phi - \frac{1}{\epsilon} \left( \frac{ \Phi'}{ a H} +  \Phi \right)  \, ,
\ea
in which a prime indicates the derivative with respect to the conformal time $\tau$ where $d \tau = dt/a(t)$.  

In general, Eq. (\ref{Eins}) can not be solved exactly because of the slow-roll correction coming from $\dot H$.
Here, we solve it to leading order in $\epsilon = -\dot H/H^2$. Note that because of the $1/\epsilon$ factor in Eq. (\ref{R-Phi2}), we need to solve Eq. (\ref{Eins}) to first order in $\epsilon$ to find the sub-leading  corrections in $\calR$.
 
At zeroth order in $\epsilon$   and taking $a H = -1/\tau$, 
Eq. (\ref{Eins})  is cast into the simple form
\ba
\label{phi}
\Phi ''-\frac{6}{\tau}\Phi'+\frac{6}{\tau^2}\Phi+k^2 \Phi=0  \, .
\ea
The general solution is represented in terms of two independent solutions $\Phi_1^{(0)}$ and $\Phi_2^{(0)}$ in which 
\ba
\label{sol2}
&\Phi_1^{(0)}(k,\tau)=k \tau(k^2\tau^2-3)\sin k\tau +3k^2\tau^2 \cos (k\tau)\\
\label{sol1}
&\Phi_2^{(0)}(k,\tau)=k^6 \tau (k^2\tau^2-3)\cos (k\tau)-3 k^7\tau^2 \sin (k\tau) \, .
\ea
Note that the superscript $(0)$ above  indicates that we have calculated $\Phi$ to zeroth order of $\epsilon$. 
Note also the overall power of $k$ which is different for $\Phi_1^{(0)}(k,\tau)$ and $\Phi_2^{(0)}(k,\tau)$.
This is chosen for convenience in follow up calculations, as an overall power of $k$ can be absorbed into constants of integration  $C_1(k)$ and $C_2(k)$ as we shall see below. However, it is important to note that for  
each $i=1, 2$ it is the relative $k$-dependence of $\Phi_i$ and $\calR_i$ (obtained from $\Phi_i$ below) which
matters.

Having calculated the zeroth order solution of Eq. (\ref{Eins}) now we calculate the next leading term 
$\Phi_i^{(1)}(k,\tau)$ for both modes $i=1, 2$. For this we also should take into account that to next slow-roll correction in non-attractor model we have
$a H \simeq - (1+ \epsilon/7) \tau^{-1}$. The corresponding differential equation for $\Phi_i^{(1)}(k,\tau)$ obtained from perturbing Eq. (\ref{Eins})  is
\ba
\Phi_i^{(1)''} -\frac{6}{\tau}\Phi_i^{(1)'}+\frac{6}{\tau^2}\Phi_i^{(1)}+k^2 \Phi_i^{(1)}=
\frac{6 \epsilon}{7\tau}  \Phi_i^{(0)'}  + \frac{2 \epsilon}{7 \tau^2} \Phi_i^{(0)} \quad  \quad i=1, 2 \, .
\ea
The above equation for $i=1,2$ can be solved separately yielding 
\ba
\label{Phi1-epsilon}
\Phi_1^{(1)} =  \frac{\epsilon}{28  } \Big[ \cos (k \tau) \left( 21 + 4  k^2 \tau^2 \right) +  k \tau \sin (k \tau) 
\left( 5 + 2 k^2 \tau^2  \right)  \Big]  \, ,
\ea
and
\ba
\label{Phi2-epsilon}
\Phi_2^{(1)} =  \frac{\epsilon}{28  \tau^5} \Big[ \cos (k \tau) \left( -945 + 315 k^2 \tau^2  + 5 k^6 \tau^6 + 2 k^8 \tau^8   \right)  \nonumber \\
~~~~~~~~~~~~~~~~~~~ -   k \tau \sin (k \tau) \left( 945 + 21 k^4 \tau^4  +4 k^6 \tau^6 \right) \Big] 
\ea

Having calculated 
$\Phi_i= \Phi_i^{(0)} + \Phi_i^{(1)} $ we can calculate $\calR$ from Eq. (\ref{R-Phi2}),  yielding to leading order 
\ba
\label{R-Phi-1}
\calR_{1} &= &\frac{k^3 \tau^3}{\epsilon }  \Big( -\sin (k \tau) + k \tau \cos (k \tau)  \Big)  \nonumber\\
&+& \frac{1}{28 } \Big[ \cos (k \tau) ( 105  -63  k^2 \tau^2 - 2  k^4 \tau^4 )  +k \tau  \sin (k \tau) ( 105 - 16  k^2 \tau^2)   \Big]  + {\cal O} (\epsilon) \, ,
\ea
and
\ba
\label{R-Phi-2}
\calR_{2} &= &-\frac{k^8 \tau^3}{\epsilon }  \Big( k \tau \sin (k \tau) +  \cos (k \tau)  \Big) 
-  \frac{k^2 \cos (k \tau)}{28 \tau^3 }  \Big(   315  - {105} k^4 \tau^4 + 16 k^6  \tau^6  \Big)
 \nonumber\\
&-& \frac{ k^2 \sin (k \tau) k \tau}{28 \tau^3 }  \Big(  315  + 105 k^2 \tau^2 - {63} k^4 \tau^4 -{2} k^6 \tau^6  
\Big) + {\cal O} (\epsilon) \, .
\ea
Note that the general solution for $\calR$ is given in terms of two independent solutions $\calR_1$ and $\calR_2$ via
$\calR = C_1(k) \calR_1 + C_2(k) \calR_2$ in which $C_1(k)$ and $C_2(k)$ are two constants of integrations. 
As mentioned before,  $C_i(k)$ are $k$-dependent so an overall power of $k$ can be absorbed in both $\Phi_i$ and
$\calR_i$. However, for each $i$, it is the relative $k$-dependence of $\Phi_i$ and
$\calR_i$ which is important.

The above expressions for $(\Phi_1, \calR_1)$ and $(\Phi_2, \calR_2)$ are valid for both sub-horizon and super-horizon limits. Now, in order to make contact with Weinberg's theorem, 
let us look at the super-horizon limits of the above solutions corresponding to $\frac{k }{a H} = -k \tau \rightarrow 0$. In this limit  for the first mode we obtain
\ba
\label{Phi-1-super}
\Phi_1 \simeq - \frac{k^6 \tau^6}{15} + \frac{3}{4} \epsilon  
\quad  \quad   \left( k \tau  \rightarrow 0  \right) \, ,
\ea
and
\ba
\label{R-Phi-1b}
\calR_{1} \simeq   -  \frac{k^6 \tau^6}{3 \epsilon }   + \frac{15}{4}   
\quad  \quad   \left( k \tau  \rightarrow 0  \right) \, .
\ea
From the above solutions we observe that $\Phi_1 = \frac{\epsilon}{5} \calR_1$ in exact agreement
with Weinberg's theorem as given in Eq. (\ref{R-Phi-ratio}).  Also note that  in the mathematical limit $k =0$ we see that  $\calR_1$ becomes constant  as was expected. However, as we discussed in previous sub-section, 
we have to be careful when taking the 
super-horizon limit  $k \tau \rightarrow 0$ while $k$ is held fixed. In this limit 
$\epsilon \propto \tau^6$ so the first term in Eq. (\ref{R-Phi-1b}) is a constant too.
To compare the two contributions in Eq. (\ref{R-Phi-1b}), let us parameterize $\epsilon$ as
\ba
\label{epsilon-form}
\epsilon(\tau) = \epsilon_*  \left( \frac{\tau}{\tau_*} \right)^6 \, ,
\ea
in which $\tau_*$ indicates the time when the mode $k$ leaves the horizon corresponding to $k \tau_* =-1$. Plugging this 
in Eq. (\ref{R-Phi-1b}) we obtain
\ba
\calR_1 \simeq -\frac{1}{3 \epsilon_*}  + \frac{15}{4}  
\quad  \quad   \left( k \tau  \rightarrow 0  \right) \, .
\ea
From this expression we see that the first term in Eq. (\ref{R-Phi-1b}) typically dominates over the second term.

Now let us look at the second mode in super-horizon limit in which we obtain
\ba
\label{Phi-2-super}
\Phi_2 \simeq  {-}\frac{135 \epsilon}{4 \tau^5}  \left(  1+ {\frac{1}{6}} k^2 \tau^2 \right) 
\quad  \quad   \left( k \tau  \rightarrow 0  \right) \, ,
\ea
and
\ba
\label{R-Phi-2b}
\calR_{2} \simeq   -  \frac{k^8 \tau^3}{ \epsilon }   - \frac{45 k^2}{4 \tau^3 }      
\quad  \quad   \left( k \tau  \rightarrow 0  \right) \, .
\ea
In the mathematical limit $k=0$, from the above solutions we find $\calR=0$ while $\Phi_2 \propto \epsilon/\tau^5 
\propto H/a$ in agreement with the  findings of \cite{Weinberg:2008zzc, Weinberg:2003sw} for the  second mode. However, in the physical super-horizon limit in which $k \tau \rightarrow0$ while $k$ is held fixed, and with $\epsilon$ given in Eq. (\ref{epsilon-form}),  
we  obtain
\ba
\label{R-Phi-2c}
\calR_{2} \simeq - \frac{k^2}{\tau^3} \left( \frac{1}{\epsilon_*} + \frac{45}{4} \right) \, .
 \ea 
The above result indicates the $1/\tau^3$ growth of $\calR$ on super-horizon as observed in \cite{Namjoo:2012aa}.
Note that the $1/\tau^3$ growth in $\calR_{2}$ is specific to non-attractor model in which $\epsilon$ falls off exponentially. 

Now we can see how the non-attractor solution evades Weinberg's theorem. As just mentioned above, our results in the mathematical limit
$k=0$ agree with the second mode of Weinberg. However,  the physical super-horizon limit is when $k \tau \rightarrow 0$ for a {\it given } $k$. In this limit, and very similar to discussions after Eq. (\ref{Poisson2}),  
the singular $1/\tau^3$ pre-factor accompanying  $k^2$ in  $\calR_{2}$ determines the structure of the physical solution.  As we argued before,  the mathematical super-horizon limit $k \rightarrow 0$ employed in  \cite{Weinberg:2008zzc, Weinberg:2003sw}, without taking into account the strong time-dependence of $\epsilon$,  
can not capture this solution.

\subsection{From comoving gauge to Newtonian gauge }
\label{comoving-method}

In this sub-section we present the equations in comoving gauge which is more convenient for models containing
scalar fields. Then we move from  comoving gauge to Newtonian gauge which provides us with  yet another insight as how the theorem in \cite{Weinberg:2008zzc, Weinberg:2003sw} is violated.

Let us start with the ADM formalism in comoving gauge $\delta \phi=0$, 
in which the metric perturbations has the following form 
\begin{equation}
ds^2=-N^2dt^2+g_{ij}(N^idt+dx^i)(N^jdt+dx^j).
\end{equation}
Here $N$ and $N^i$ are the lapse function and the shift vectors which are obtained algebraically from the 
constraint equations. 

In comoving gauge, the spatial metric take the following simple form (neglecting transverse and traceless part)
\begin{equation}
g_{ij}=a^2(1+2\calR ) \delta _{ij}.
\end{equation}
As usual,   we may write down the quadratic action and solve for the lapse function and the shift vector. 
Defining the lapse function and the shift vector via \cite{Maldacena:2002vr}
\begin{equation}
\label{Ni}
g_{0i}=N_i \equiv \partial _i \psi, \qquad g_{00} \equiv -(1+2N_1),
\end{equation}
from the constraint equations we obtain 
\begin{equation}
N_1=\frac{2\dot{\calR}}{H},
\end{equation}
and,
\begin{equation}
\label{psi-sol}
\psi = -\frac{\calR}{H}+\chi , \qquad \chi \equiv  \partial ^{-2}(a^2\epsilon \dot{\calR}).
\end{equation}

Note that in usual attractor case in which $\calR$ is conserved outside horizon we have
\begin{equation}
\dot{\calR}\sim \frac{k^2}{a^2H} \calR ,
\end{equation}
so  $\chi$ is analytic in $k$. However, in non-attractor case in which  \cite{Namjoo:2012aa}
\begin{equation}
\dot{\calR}=-3H\calR +\mathcal{O}(k^2/a^2H^2),
\end{equation}
then $\chi$ is non-analytic in $k$. This is another sign that the prescription of taking $k \rightarrow 0$ employed
in  \cite{Weinberg:2008zzc, Weinberg:2003sw} as the definition of super-horizon limit is problematic. 

Now we perform the coordinate transformation from the comoving gauge to the Newtonian gauge. 
 Consider the coordinate transformation 
 \begin{equation}
x^i\rightarrow x^i+\xi ^i, \qquad \xi ^i = \partial _i \epsilon ^S \, ,
\end{equation}
in which $\epsilon ^S$ is the scalar part of spatial coordinate transformation. 

If we split the metric as $g_{\mu \nu}=\overline{g}_{\mu \nu}+h_{\mu \nu}$,  under the above coordinate transformation  we have,
\begin{align}
\label{Delta-hi0}
\Delta h_{i0}&=\partial _i \left(-\dot{\epsilon}^S-\epsilon ^0 +2H\epsilon ^S\right) , \\
\label{Delta-hij}
\Delta h_{ij}&=-2\partial _i \partial _j \epsilon ^S + 2a^2H \epsilon ^0 \delta _{ij}, \\
\label{Delta-h00}
\Delta h_{00}&=-2\dot{\epsilon}^0,
\end{align}
in which $\Delta h_{\mu \nu}$ indicates the change in $h_{\mu \nu}$ in transforming from the comoving
gauge to the Newtonian gauge.

In the Newtonian gauge we should keep the spatial metric diagonal so  from Eq. (\ref{Delta-hij}) we require\footnote{Note that in general $\epsilon ^S=f(t)$ will keep the spatial metric diagonal too. However, this choice gives rise to pure gauge mode which has been already taken care of in Weinberg's theorem.}
\begin{equation}
\epsilon ^S=0.
\end{equation}
In addition, in Newtonian gauge $h_{0i}=0$ and taking into account that in comoving gauge $h_{0i}= N_i$ is given in Eqs.  (\ref{Ni}) (\ref{psi-sol}), from Eq. (\ref{Delta-hi0}) we obtain
\begin{equation}
\partial _i \left[ -\frac{\calR}{H}+\chi - \epsilon ^0\right] =0 \, .
\end{equation}
Therefore, neglecting pure gauge mode, from this equation  we obtain
\begin{equation}
\epsilon ^0 = -\frac{\calR}{H}+\chi = -\frac{\calR}{H}+\partial ^{-2}\left(a^2\epsilon \dot{\calR} \right).
\end{equation}

Now, plugging this value of $\epsilon^0$ into Eqs. (\ref{Delta-h00}) and (\ref{Delta-hij}) the components of metric in Newtonian gauge is obtained to be 
\ba
\label{g00-Newtonian}
g_{00}&=&-1-\frac{2\dot{\calR}}{H}-2\dot{\epsilon}^0=-1+2\epsilon \calR - 2 \partial _t \partial ^{-2}\left( a^2\epsilon \dot{\calR}\right), \\
\label{gij-Newtonian}
g_{ij}&=&a^2\left[1+2H\partial ^{-2}\left( a^2\epsilon \dot{\calR}\right) \right] \delta_{ij} \, .
\ea
The above expressions for $g_{00}$ and $g_{ij}$ give two independent formulas for $\Phi$ and $\Psi$.
Now imposing the constraint $\Phi=\Psi$ in Newtonian gauge, we readily obtain the second order differential
equation for $\calR$ as given in Eq. (\ref{R-eq}). In addition, once $\calR$ is solved this way, we can plug
it into Eq. (\ref{gij-Newtonian}) to obtain $\Phi$ as follows 
\begin{equation}
\label{Phi-Newtonian}
\Phi =\Psi = -H \partial ^{-2} \left(a^2 \epsilon \dot{\calR} \right)=-H\chi.
\end{equation}
Note that the above solution works for both attractor and non-attractor phases, and it is physical  because we obtained it from coordinate transformation of a physical solution in comoving gauge. 

Now, as it is  stressed earlier, in attractor case $\chi$ is analytic in $k$ i.e. it is well defined in $k\rightarrow 0$ limit.  Therefore both of Weinberg's adiabatic modes are physical and the theorem works well. This is also seen from the explicit solutions of $\calR$ in Eq. (\ref{R-eq}) as discussed in previous sub-section. However, in the non-attractor  case that $\calR$  evolves on super-horizon scales  $\chi$ is non-analytic in $k$ so the limit $k\rightarrow 0$ is not well defined mathematically. This is also seen from the structure of Eq. (\ref{Phi-Newtonian}) in which 
$\Phi= (a^2 \epsilon/k^2) \dot \calR$. The analyticity of the results for the limit $k\rightarrow 0$ requires that 
$\dot \calR =0$. Conversely, if we do not know $\dot \calR =0$ {\it  a priori } then we can not assume the analyticity of the solutions
in the limit $k\rightarrow 0$ which is taken as the guiding principle to distinguish the physical solution from the pure gauge mode.  

\section{Fluid inflation}
\label{fluid}

Fluid inflation, presented originally in  \cite{Chen:2013kta}, is another example in which Weinberg's theorem is violated.  Here we briefly review the setup of fluid inflation and present the reasons why  it violates 
Weinberg's theorem in close analogy with non-attractor scenarios. 

The fluid setup is given by the following Lagrangian density \cite{Ray1, Ray2}
\ba
{\cal L}= \dfrac{1}{2} M_P^2 \sqrt{-g} R - \sqrt{-g} \, {\rho} (1+ e(\rho)) + \sqrt{-g} \lambda_1 \left(  g_{\mu \nu} U^{\mu} U^\nu +1 \right) + \sqrt{-g} \, \lambda_2 \left( \rho U^\mu   \right)_{;\mu}\,,
\ea
in which $M_P$ is the reduced Planck mass, 
$\rho$ is the rest mass density, $e(\rho)$ is the
specific internal energy and  $U^{\mu}$ is the 4-velocity. In addition,  $\lambda_1$ and $\lambda_2$ are two Lagrange multipliers to enforce the  normalization of the 4-velocity and the  conservation of the rest mass density.
With this prescription, the total energy density, $E$, is given by
\ba
\label{E-eq}
E = \rho (1+e)\,.
\ea
As in  \cite{Chen:2013kta}  we concentrate on an isentropic or barotropic fluid for which $e= e(\rho)$. Having this said, there is no restriction to  consider more general situations in which $e$ can also be  a function of other thermodynamic variables such as entropy.

Varying the action with respect to the Lagrange multipliers  $\lambda_1$ and $\lambda_2$ and dynamical fields 
$\rho$ and $g_{\mu \nu}$ we recover the Einstein's fields equation in which now the stress energy tensor $T^{\mu \nu}$ takes the form of a perfect fluid
 \ba
T^{\mu \nu}=(E + P) U^\mu U^\nu+ P g^{\mu \nu}\, .
\ea
Here $P$ plays the role of pressure in which for an isentropic fluid is represented by
\ba
\label{P}
\dfrac{d e (\rho)}{d\rho}= \dfrac{P}{\rho^2}\,.
\ea
Knowing that $e=e(\rho)$, from the above equation we conclude that $P$
is a function of $\rho$. Alternatively, from  Eq.~(\ref{E-eq}) we also conclude
\ba
\label{E-diff}
\frac{d E}{d \rho} = \frac{E+P}{\rho} \,.
\ea
We note that Eqs. (\ref{P}) and (\ref{E-diff}) imply that $P$ is a function of $E$,
$P=P(E)$, which is expected for a barotropic fluid.

An important parameter of the fluid is the sound speed of perturbations $c_s$ which is given by
\ba
\label{cs-def}
c_s^2 \equiv \frac{\dot P}{\dot E}\, .
\ea
For a small perturbation, and using  the conservation equation $\dot E + 3 H (E + P)=0$,   this implies
\ba
\label{P-rho}
\delta P = c_s^2 \delta E = c_s^2 (E+ P) \frac{\delta \rho}{\rho}\,.
\ea
Note that the definition (\ref{cs-def}) makes sense as  we consider a barotropic fluid. In order for the perturbations to be stable we require $c_s^2 >0$, while for the perturbations to be sub-luminal we also require $c_s^2 \leq 1$.

The  cosmological dynamics of the system has the usual FRW form. However, as compared to inflation based on  scalar field dynamics,  we note that for fluid  setup the total energy density $E$ and the pressure $P$ internally are functions of the rest mass density $\rho$. This yields a non-trivial equation of state  $P= P(E)$   for a barotropic fluid
in which $c_s$ plays non-trivial roles in perturbation analysis. 

The first and second slow-roll parameters $\epsilon$ and $\eta$ respectively are
\ba
\epsilon = -\frac{\dot H}{H^2} = \frac{E +P}{2 M_P^2 H^2}\, ,
\ea
and
\ba
\label{eta-eq}
\eta \equiv \frac{\dot \epsilon}{H \epsilon} = 2 \epsilon - 3 (1 + c_s^2) \, .
\ea
From the form of $\eta$ we see the important difference compared to conventional slow-roll models of scalar field theories. Requiring that $0 < c_s^2 \leq 1$, and taking $\epsilon \ll 1$ in order to sustain a long enough period 
of inflation, we conclude that $-6 \lesssim \eta \lesssim -3$.  At this stage we can not pin down the exact value of
$\eta$, this should be fixed from the scale-invariance of the  curvature perturbation power spectrum. However, for $\eta$ given in the above range, we readily conclude that $\epsilon$ falls off exponentially which, as we shall see below,  closely resembles  the non-attractor scenario. 

To Perform the cosmological perturbation analysis we go to comoving gauge defined on a time-slicing  in which the fluid's 4-velocity is orthogonal to the hypersurface $t= \mathrm{constant}$  and the three-dimensional spatial metric is conformally flat \cite{Chen:2013kta}. Calculating the quadratic action  in comoving gauge we obtain 
\ba
\label{R-eq-fluid}
(z^2 \calR')' + c_s^2 k^2 z^2 \calR =0\, ,
\ea
in which a prime denotes the derivative with respect to conformal time and $z$ is defined via $z^2= 2 \epsilon a^2/c_s^2$. We note that the above equation for $\calR$ is similar to Eq. (\ref{R-eq}) obtained for scalar field theory. 
Now quantizing the system and calculating the power spectrum, the spectral index is obtained to be $n_s \simeq 3 (1- c_s^2)$  \cite{Chen:2013kta}.   We see that to obtain a scale invariant power spectrum we require $c_s^2 =1$. Consequently, from
Eq. (\ref{eta-eq}) we conclude that $\eta \simeq -6$ and hence $\epsilon \propto a(t)^{-6}$. Very interestingly, we see that fluid inflation is a non-trivial realization of non-attractor setup,  completely independent of scalar field dynamics.  Now it should not be surprising that in fluid setup, $\calR$ is not frozen on super-horizon scales and indeed we readily conclude that $\calR \propto a(t)^{3}$  \cite{Chen:2013kta}.

Having established the direct link between the fluid setup and the non-attractor setup, we can use any of the arguments presented in sub-sections \ref{R-method}, \ref{Phi-method} or  \ref{comoving-method}  to understand why Weinberg's theorem is violated in  the model of fluid inflation. For example in the method of sub-section \ref{R-method},  in Poisson constraint Eq. (\ref{Poisson2}) we find that $\epsilon$ falls off much stronger than the combination  $k^2 /a^2 H^2$ so  one can not  take $k\rightarrow 0$ arbitrarily for a given $k$ to define the super-horizon limit. Or in the method of subsection  \ref{Phi-method}, with $c_s=1$
we conclude that $\delta P= \delta E$, and similar to non-attractor case,  Eq. (\ref{Eins}) can be solved directly to 
find $\Phi$. The rest of the argument as how the  theorem is violated in fluid inflation setup 
goes parallel to the discussions after Eq. (\ref{R-Phi-2c}).

\section{Solid inflation }
\label{solid}

In this section we  study  the model of solid inflation  \cite{Endlich:2012pz} which is another known example in literature which  violates Weinberg's theorem;  for other works on solid inflation see 
\cite{Endlich:2013jia, Bartolo:2013msa, Akhshik:2014bla, Akhshik:2014gja, Bartolo:2014xfa, Dimastrogiovanni:2014ina}. 

As the name indicates, in this model   inflation is driven by 
a configuration resembling a solid. In this setup the three-dimensional space is divided into small cells 
such that  the location of each cell is defined by the value of scalar fields $\phi^I$ for $I=1, 2$ and $3$. More specifically, at the background level the position of each cell is represented by
\ba
\label{back-phi}
\langle \phi^I \rangle = x^I   \quad , \quad I=1, 2, 3.
\ea
At this stage the ansatz (\ref{back-phi}) naively seems to violate the isotropy and the homogeneity of the cosmological background as the scalar fields $\phi^I$ are time-independent and depend explicitly on $x^I$. 
However, on the physical grounds,  one should impose the following internal symmetries to  keep the background isotropic and homogeneous
\ba
\label{cond1}
\phi^{I} \rightarrow \phi^I + C^I \, ,
\ea
and
\ba
\label{cond2}
\phi^I \rightarrow O^I_J \phi^J \quad , \quad O^I_J \in SO(3) \, .
\ea
We note that  $C^I$ are constants while $O^I_J$ belong to $SO(3)$ rotation group. The symmetry under  translation  in field space imposed by Eq. (\ref{cond1}) enforces that the dynamical quantities in the Lagrangian are constructed from derivatives of the scalar fields $\partial \phi^I$. Consequently,  the background  Eq. (\ref{cond1}) becomes invariant under translation. Furthermore, the internal $SO(3)$ rotation invariance guarantees  the isotropy of the background. In conclusion,  with the  internal symmetries (\ref{cond1}) and (\ref{cond2}) enforced, the background  is consistent with the cosmological principles. 

The most general action consistent with the above internal symmetries which is  minimally coupled to gravity  
 is given by 
\ba
\label{solid-action}
S = \int d^4 x \sqrt{-g} \left\{ \frac{M_P^2}{2} R + F[X, Y, Z] 
\right\} \, ,
\ea
 in which  $M_P$ is the reduced Planck mass related to Newton constant via $M_P^2 =1/8 \pi G$ and 
 $F$ is a function incorporating the properties of the solid.  The condition that the action is invariant under the internal  symmetries (\ref{cond1}) and (\ref{cond2}) requires that the variables $X, Y$ and $Z$ are functions of the derivatives of $\phi^I$ which in turn are given in terms of   the $SO(3)$ invariant matrix $B^{IJ}$
 via 
 \ba
\label{XYZ-def}
X \equiv  [B]  \quad , \quad
Y \equiv \frac{[B^2]}{ [B]^2} \quad , \quad
Z\equiv \frac{ [B^3]}{  [B]^3}  \, ,
\ea
in which $[ B]  \equiv  \mathrm{Tr}(B) $ and 
  \ba
 \label{B-def}
 B^{IJ} \equiv g^{\mu \nu} \partial_\mu \phi^I \partial_\nu \phi^J \, .
 \ea
 Our convention is that the Greek indices $\mu, \nu, ...$ indicate the four-dimensional spatial coordinates while the capital Latin indices $I, J, ...$ represent the three-dimensional  internal matter field space.

At the background level we can check that 
\ba
X= \frac{3}{a(t)^2} \quad , \quad Y= \frac{1}{3}  \quad , \quad Z=\frac{1}{9} \, .
\ea
Note that the variables $Y$ and $Z$ are constructed such that they  are insensitive to the volume of 3-space while the information about the background volume is entirely encoded in $X$. 

The energy momentum-tensor  is given by 
\begin{equation}
\label{T-munu}
T^\mu_\nu=\delta ^\mu _\nu F-2g^{\mu \alpha}\partial _\alpha \phi ^I \partial _\nu \phi ^J M^{IJ} \,  ,
\end{equation}
in which  we have defined $M^{IJ}$ via
\begin{equation}
\label{MIJ}
M^{IJ}\equiv \left(F_X-\frac{2F_YY}{X}-\frac{3F_ZZ}{X}\right)\delta ^{IJ}+\frac{2F_YB^{IJ}}{X^2}+\frac{3F_ZB^{IK}B^{KJ}}{X^3} \, ,
\end{equation}
where $F_X \equiv \partial F/\partial X$ and so on. 

With the above form of $T^{\mu}_{ \nu}$, the energy density $\rho$ and the pressure $P$ at the background level are given by
\ba
\label{rho-p}
\rho = -F \quad , \quad P= F- \frac{2}{a^2} F_X \, ,
\ea
yielding the expected  cosmological equations 
\ba
\label{back-eq}
3 M_P^2 H^2 = \rho \quad , \quad  \dot H = -\frac{1}{2 M_P^2} (\rho + P) \, .
\ea
On the other hand,  by varying the action with respect to $\phi^I$,  the  scalar fields equations is obtained to be
\ba
\label{KG-eq}
\partial_\mu \left( \sqrt{-g} \frac{\partial F}{\partial B^{ab}} \frac{\partial B^{ab}}{\partial \partial_\mu \phi^I} \right) =0  \, .
\ea
We note the curious effect that at the background level $\phi^I$ are independent of $t$
and Eq. (\ref{KG-eq})  is automatically satisfied so we do not get any information from Eq. (\ref{KG-eq}) at
the background level.


At this level it may look that the solid scenario is a model with three inflationary fields $\phi^I$ which can generate entropy perturbations which can naturally bypass Weinberg's theorem. However, as studied in  \cite{Endlich:2012pz},  the  scalar  perturbations are generated  effectively by one degree of freedom.  This scalar perturbation is described by the single field $\pi_L$ corresponding to the longitudinal component of  of the fluid excitations, which are  dubbed as ``phonons" in \cite{Endlich:2012pz}.  More specifically,  suppose 
\ba
\phi^I = x^I + \pi^I(t, \bfx) \,  ,
\ea
and decompose the filed $\pi^I$ into its transverse and longitudinal parts as 
\ba
\label{pi-decompose}
\pi^i(t, \bfx) = \frac{{\partial_i}}{\sqrt{-\nabla^2}} \pi_L(t, \bfx) + \pi^i_T(t, \bfx) \, ,
\ea 
in which $\partial_i \pi^i_T =0$.  In this decomposition, $\pi_L$ sources the  curvature perturbations while 
$\pi_T^i$ sources the vector perturbations. Note that  we do not pursue  the $\pi_T^i$ excitations any further   because the vector perturbations are damped after  inflation. 

Going to flat gauge, the curvature perturbations is given by $\zeta = -\frac{k}{3} \pi_L$ which  on super-horizon scales is obtained to be
\ba
\label{zeta-super}
\zeta(\tau) \propto (-c_L k \tau)^{-A} \left(  1+ B \ln  (-c_L k \tau) \right) \, ,
\ea
in which $A$ and $B$ are constants of order slow roll parameters $\epsilon$ and $c_L\simeq 1/\sqrt3$ is the sound speed of phonons.  From the above expression we  observe a mild running of curvature perturbation on super-horizon scales  varying like $\epsilon N$ in which $N= \ln(- k \tau)$ is the number of
e-folds before the end of inflation. Our goal in this Section  is to understand how this happens, bypassing the theorem in \cite{Weinberg:2008zzc, Weinberg:2003sw}.

To address this question,  we obtain the perturbed Einstein equations in Newtonian gauge.  For this purpose, first we need the components of  the perturbed  energy momentum tensor $\delta T^\mu_\nu$. Using Eq. (\ref{T-munu}) we have
\ba
\label{delta-T-00a}
\delta T^\tau_\tau = \delta F = F_X \delta X + F_Y \delta Y + F_Z \delta Z \, .
 \ea
However, with some efforts one can show that $\delta Y$ and $\delta Z$ vanish up to linear oder in perturbations so we can neglect their contributions and  $\delta T^\tau_\tau = F_X \delta X$. 
On the other hand for $\delta X$ we have
\ba
\label{delta-X}
\delta X &=& \delta g^{ii} + \frac{2}{a^2} \partial_i \delta \phi^i  \nonumber\\
&=& \frac{2 X}{3} ( 3 \Psi - k \pi_L) \, ,
 \ea 
in which  the relation $\pi^i = \frac{1}{k} \partial_i \pi_L$ has been used.  As a result, for $\delta T^\tau_\tau$ component  we obtain
\ba
\label{delta-T-00}
\delta T^\tau_\tau = 2 X F_X ( \Psi - \frac{k}{3} \pi_L) \, .
\ea
Similarly, for $\delta T^\tau_i$ component we have
\ba
\label{delta-T-0i}
\delta T^\tau_i = \frac{2 X}{3} F_X  \delta {\phi^i}^\prime \, ,
\ea
in which a prime indicates the derivative with respect to conformal time $\tau$. 

On the other hand, the calculation of $\delta T^i_j$ is more non-trivial. We have
\ba
\label{delta-T-ija}
\delta T^i_j = F_X \delta X \delta^i_j - \frac{4}{a^2} \Psi M_{ij} - \frac{2}{a^2} F_X \Pi^{ij} - \frac{2}{a^2} \delta M^{ij} \, ,
\ea
in which $\Pi^{ij}$ is  defined via
\ba
\label{Pi-ij}
\Pi^{ij} \equiv \partial_i \pi^j + \partial_j \pi^i \, .
\ea
On the other hand, one can show that
\ba
\delta M^{ij} = F_{XX} \delta X \delta^{ij} -  \frac{2( F_Y + F_Z) }{3 X^2} \delta X \delta_{ij} + \frac{2 (F_Y + F_Z)}{3 X}  \left( 2 \Psi \delta_{ij} + \Pi^{ij} \right) \, .
\ea
Plugging this expression in Eq. (\ref{delta-T-ija}) we obtain
\ba
\label{delta-T-ij}
\delta T^i_j = 
\Big( F_X - \frac{2X}{3} F_{XX} \Big) \delta X \delta^i_j - \frac{4   X F_X \Psi}{3}  \delta^i_j - \frac{2X}{3}  F_X \Pi^{ij}
+ \frac{4}{9}  ( F_Y + F_Z) \Big[ ( \frac{\delta X}{X}- 2 \Psi) \delta^i_j  - \Pi^{ij}  
\Big] 
\ea

So far no assumption was made beyond the linear perturbation theory. To simplify the analysis we impose the slow-roll assumptions and ignore terms higher in powers of the slow-roll parameter $\epsilon$. To leading order in $\epsilon$ one can show that  $c_L^2 \simeq \frac{1}{3}$, $F_Y \simeq - F_Z$ and $F_{XX} \simeq -\frac{F_X}{X}$  \cite{Endlich:2012pz}.   Putting the above results  together we obtain the following set of perturbed Einstein equations 
\ba
\label{i-neq-j}
\Phi -\Psi  &= &\frac{4  \left(\mathcal{H}^\prime -\mathcal{H}^2\right) }{k} \pi_L \\
\label{0i-eq}
\Psi ^\prime + \mathcal{H} \Phi    &=&  -  \frac{    \left(\mathcal{H}^\prime-\mathcal{H}^2\right) }{k} \pi_L'  \\
\label{Ein3}
12  \mathcal{H} \Psi' + 3 k^2 \Psi + ( k^2 + 12 \mathcal{H}^2 ) \Phi & =& 12   \left(\mathcal{H}^\prime-\mathcal{H}^2\right) \Psi \\
\label{Ein4}
\Psi'' + k^2 \Psi + 5   \mathcal{H} \Psi' +  \mathcal{H} \Phi' + (  2  \mathcal{H}' + 4  \mathcal{H}^2 ) \Phi &=&
 \left(\mathcal{H}^\prime-\mathcal{H}^2\right)  \left( 6 \Psi - \frac{2}{3} k \pi_L \right) \, .
\ea
in which $\calH \equiv  a H$.

Note the interesting conclusion  from Eq. (\ref{i-neq-j})  that, unlike conventional models of inflation,
$\Psi \neq \Phi$.  This is because in the model of solid inflation the longitudinal mode $\pi_L$ sources the anisotropic stress $\pi^S$ and therefore we have $\phi \neq \Psi$. To see this explicitly, note that $\pi^S$ is related to
 $\delta T^i_j $ via  $\delta T^i_j = \delta P \delta^i_j + \partial_i \partial_j \pi^S$ \cite{Weinberg:2008zzc}. Now
 with $\delta T^i_j $ given in Eq. (\ref{delta-T-ij}), in the slow-roll limit, we obtain 
\ba
\label{Pi-S}
\pi^S = -\frac{4 \epsilon F}{3 k}  \pi_L \, .
\ea
On the other hand,  the $i \neq j$ component of the perturbed Einstein equation in general is written as  \cite{Weinberg:2008zzc}
\ba
\partial_i \partial_j ( \Phi - \Psi) = -\frac{a^2}{M_P^2} \partial_i \partial_j \pi^S \, .
\ea
Now with the form of $\pi^S$ given in Eq. (\ref{Pi-S}) we obtain Eq. (\ref{i-neq-j}). Also note that 
Eq. (\ref{Pi-S}) shows the $1/k$ non-analytic relation  between $\pi^S$ and $\pi_L$
which directly violates the analyticity assumption employed
in the proof \cite{Weinberg:2008zzc, Weinberg:2003sw}. Consequently, it should not be surprising that 
the conclusion in \cite{Weinberg:2008zzc, Weinberg:2003sw} is violated in solid inflation.

As another sign of non-analytic structure of solid model, note that  Eq. (\ref{0i-eq}) represents the momentum conservation equation, i.e. the $(0i)$ component of Einstein equation,  
in which the scalar velocity potential (in convention of  \cite{Weinberg:2008zzc})  is obtained to be 
\ba
\label{u-solid}
\delta u = -\frac{a}{k} \pi_L' \, .
\ea
Again, we see the non-analytic $1/k$ behavior in fields' equations  as discussed above. 

One can  eliminate $\pi_L$ and $\Phi$ in favors of $\Psi$ and obtain a closed second order differential equation  for $\Psi$.  For this purpose from Eqs. (\ref{i-neq-j}) and (\ref{Ein3}) we obtain
\ba
\label{piL-Psi}
\pi_L &=& \frac{k \Big[  ( 3 \mathcal{H}^\prime - 6 \mathcal{H}^2 - k^2) \Psi - 3  \mathcal{H} \Psi'     \Big]  }{ \left(\mathcal{H}^\prime-\mathcal{H}^2\right) ( k^2 + 12 \mathcal{H}^2) }  \, ,\\
\label{Phi-Psi}
\Phi &=& \frac{ 12 \left(\mathcal{H}^\prime-\mathcal{H}^2\right) \Psi - 3 k^2 \Psi - 12  \mathcal{H} \Psi'}{k^2 + 12 \mathcal{H}^2} \, .
\ea
Now plugging the above expressions for  $\pi_L$ and $\Phi$ in   Eq. (\ref{0i-eq}), and using the following 
relations which is valid in slow-roll limit
\ba
\mathcal{H}^\prime-\mathcal{H}^2 \simeq \frac{\epsilon}{ \tau^2 } \quad , \quad \mathcal{H}'' \simeq 2 \mathcal{H}^2 \, ,
\ea
we obtain our desired equation for  $\Psi$
\ba
\label{Psi-eq}
3 \big( k^2 + 12 \mathcal{H}^2 \big) \Psi'' - 72 \mathcal{H}^3 \Psi' + \Big(  k^4 - 12 \mathcal{H}^2 ( k^2 + 6 \mathcal{H}^2   )   \Big) \Psi = 0 \, .
\ea
Happily Eq. (\ref{Psi-eq}) can be solved analytically. Imposing the Minkowski initial condition for the modes inside the horizon ( corresponding to $k| \tau | \gg 1$),  we obtain 
\ba
\label{Psi-sol}
\Psi (x) = -\sqrt{\frac{3}{2 k}} \frac{1}{x^2} ( 2 \sqrt3 + i x)^2 e^{-\frac{ix }{\sqrt{3}}} \, ,
\ea
in which we have defined $x \equiv k \tau$ .  Note that the factor $1/\sqrt{3}$ in the exponent appears because 
the modes deep inside the horizon propagate with the sound speed  $c_L^2 \simeq \frac{1}{3}$.  

Now let us look at the above solution in the super-horizon limit $x \rightarrow 0$
\ba
\label{Psi-sol2}
\Psi \propto \frac{1}{(k \tau)^2} = e^{2N} \quad \quad (k \tau \rightarrow 0) \, ,
\ea
in which $N$ is the number of e-fold towards the end of inflation  with the convention $N>0$. The above equation clearly demonstrates that on super-horizon scales  the gravitational potential grows exponentially.  This non-perturbative growth of $\Psi$ implies that the Newtonian gauge is not a reliable  gauge to study perturbations in solid inflation. 

Now with $\Psi$ calculated in Eq. (\ref{Psi-sol}) we can  calculate $\zeta$. Knowing that  $\zeta $ is given by  $\zeta = -\frac{k}{3} \pi_L$, from  Eq. (\ref{Psi-sol}) we can calculate $\zeta$ yielding Eq. (\ref{zeta-super}) to leading order in slow-roll corrections.

It is important to note that because of the non-zero anisotropic stress $\pi^S$,  we have $\Psi \neq \Phi$. However, this by itself is not the source of violation of the Weinberg's theorem. Instead,  the non-analytic relation between $\pi^S$ and $\pi_L$,  as given in Eq. (\ref{Pi-S}),  is the key reason for the violation of  this theorem in solid inflation.  Note that because $\zeta =-\frac{k}{3} \pi_L$,   Eq. (\ref{Pi-S}) also implies the non-analytic relation 
\ba
\label{PiS-zeta}
\pi^S   \sim \frac{\zeta}{k^2} \, .
\ea
In addition, from Eq. ( \ref{u-solid}) we also have the non-analytic relation  between $\delta u$ and  
$\zeta$. These non-analytic  behaviors between $\pi^S$, $\delta u$ and  $\zeta$ are in direct conflicts 
with the analyticity assumption employed in the proof of \cite{Weinberg:2008zzc, Weinberg:2003sw}, as also mentioned in  \cite{Endlich:2012pz} ( see also \cite{Berezhiani:2014kga, Cannone:2014uqa, Cannone:2015rra}). 

Finally we also comment that in solid model $\calR \neq -\zeta$, even on super-horizon scales. This is because $\zeta$ is not frozen on super-horizon scales yielding $\calR \simeq -c_L^2 \zeta$ on these scales.

\section{Pseudo-conformal universe}
\label{conformal-sec}

In this section we study yet another example in literature which is known to violate the theorem in  \cite{Weinberg:2008zzc, Weinberg:2003sw}, the pseudo-conformal universe. This model was proposed in
\cite{Hinterbichler:2011qk} as an alternative to inflation which relies on conformal symmetries capable of generating nearly scale invariant power spectrum while solving the flatness and the horizon problems. 
The model shares similarities to the  $U(1)$ model 
\cite{Rubakov:2009np, Libanov:2011zy,  Libanov:2011zy} 
and the  Galilean Genesis scenario  \cite{Creminelli:2010ba}.  In the model of pseudo-conformal universe it is assumed that the early universe (before the big bang) enjoys an approximate conformal symmetry in a near flat background. At this early stage one or more of the conformal fields develop time-dependent expectation values which break the conformal symmetry. In addition, it is assumed that there are other fields with zero  conformal weight (i.e. isocurvature fields) which acquire a nearly scale-invariant power spectrum generating the observed curvature perturbations.

To be specific, and following \cite{Hinterbichler:2011qk}, we consider a simple model containing the negative
quartic potential $V= -\frac{\lambda}{4} \phi^4$ with $\lambda >0$ 
which is minimally coupled to gravity. The model is classically conformal invariant. It is assumed that there are sub-leading corrections that can uplift the potential making  the potential bounded from below. One mode of $\delta \phi$ perturbations  is freezing while the other mode grows on super-horizon scales. The latter is the mode of interest which violates the theorem in  \cite{Weinberg:2008zzc, Weinberg:2003sw}. However, as noted above, the observed curvature perturbations are generated by the additional field $\chi$ which has  the conformal weight zero  and at the background level has no expectation values, 
$\chi=0$. However, we will not study this field as we are interested to see how the growing mode of the conformal field  fluctuation $\delta \phi$ violates  Weinberg's theorem. 

In the past infinity  $t =-\infty$, the scalar field starts rolling from $\phi=0$. As the scalar field develops an expectation value and the conformal invariance is broken the universe starts a slow phase of contraction 
in which gravity is very weak, corresponding to $\lambda M_P^2 t^2 \gg 1$,  
and calculations can be accurately approximated to leading orders of $1/M_P^2$. 

The leading order $1/M_P^2$ corrections to the slowly-contracting scale factor $a(t)$, the Hubble expansion 
rate $H$ and the zeroth order evolution of $\phi(t)$  were presented in \cite{Hinterbichler:2011qk}. Here, we extend these results to next leading order $1/M_P^4$ in order to consistently calculate the next order corrections in $\Phi$ and $\calR$. To order $1/M_P^4$  we have 
\begin{align}
\label{backg}
a(t)&=1-\frac{1}{6\lambda M_P^2 t^2}-\frac{13}{360\lambda^2 M_P^4 t^4}+...\\ 
H(t)&=\frac{1}{3\lambda M_P^2 t^3}+\frac{1}{5\lambda^2M_P^4 t^5}+...\\ 
\phi(t)&=\sqrt{\frac{2}{\lambda}}\frac{1}{t}+\sqrt{\frac{2}{\lambda}}\frac{1}{6\lambda M_p^2 t^3}+\frac{19}{360}\sqrt{\frac{2}{\lambda}}\frac{1}{\lambda^2 M_P^4 t^5}+... \, .
\end{align}

Note that in this model universe is in a phase of slow contraction  so modes leave the horizon smoothly similar to an  inflationary background. The criteria for the mode to be super-horizon is 
$k |t| \ll \frac{1}{\sqrt \lambda M_P  |t| } $  \cite{Hinterbichler:2011qk}. Note that $t<0$ so that is why we have used  $|t|$. On the other hand, in order for the gravitational back-reaction to be small
we require $M_P  |t| \gg1$. Combining these two conditions we have
\begin{equation}
k |t|  < \frac{1}{\sqrt \lambda M_P |t| }  \ll 1  \, .
\end{equation}

From the background solutions we can calculate $\epsilon=-\dot H/H^2  = 9 \lambda M_P^2 t^2$. From the  weak gravity condition this implies that $\epsilon \gg 1$. As we shall see below, the strong time-dependence of $\epsilon$ plays crucial roles in violating Weinberg's theorem. 

Our strategy here is very similar to the strategy employed in sub-section \ref{Phi-method}. We would like to calculate $\Phi$ to leading orders in $1/M_P^2$ and then calculate $\calR$ and see how the theorem in \cite{Weinberg:2008zzc, Weinberg:2003sw} is violated.
The corresponding equations for $\delta \phi$ and $\Phi$ are as in Eqs. (\ref{Eins}) and (\ref{0i}) in which
now ${\delta \rho-\delta P =-2\lambda \phi^3 \delta \phi}$. Using Eq. (\ref{0i})  to eliminate 
$\delta \phi$,  from  from  Eq. (\ref{Eins}) we obtain 
\begin{equation}
\ddot{\Phi}+(7-\frac{2\lambda \phi^3}{H\dot{\phi}})H\dot{\Phi}+(6H^2+2\dot{H}+\frac{k^2}{a^2}-\frac{2\lambda \phi^3}{\dot{\phi}}H)\Phi=0 \, .
\end{equation}
Plugging the background values of $a(t), H(t)$ and $\phi(t)$ into the above equation, to leading order of
$1/M_P^2$ we obtain
\ba
\label{ode}
&\ddot{\Phi}_k+\left(\frac{4}{t}+\frac{7}{3\lambda M_P^2 t^3} \right)\dot{\Phi}_k 
&+ \left( k^2+\frac{k^2}{3\lambda M_P^2 t^2}-\frac{2}{3}\frac{1}{\lambda M_P^2 t^4} \right)\Phi_k=0 \, .
\ea

Now we solve Eq. (\ref{ode}) order by order in powers of $1/M_P^2$. At the zeroth order the solutions are given by 
\ba 
{\Phi}_1^{(0)} &=&\frac{1}{t^3} \Big( (kt)\cos (kt)-\sin (kt) \Big)\\ 
{\Phi}_2^{(0)} &=&\frac{1}{t^3} \Big(\cos (kt)+(kt) \sin (kt) \Big) \, .
\ea
Now if we take the mathematical limit $k \rightarrow 0$ it is easy to check that
\ba
{\Phi}_1^{(0)} \rightarrow  -\frac{k^3}{3} \quad , \quad \calR_1^{(0)} \rightarrow \frac{k^3}{3}\, ,
\ea 
and
\ba
{\Phi}_2^{(0)} \rightarrow  \frac{1}{t^3} \quad , \quad \calR_2^{(0)} \rightarrow -\frac{{k^2}}{3 t} \, .
\ea 
In particular, the above expressions yields ${\Phi}_1^{(0)} = -\calR_1^{(0)}$ in agreement with 
Eq. (\ref{set1}) while ${\Phi}_2^{(0)} \propto \frac{H}{a}$ and $\calR=0$ in agreement with Eq. (\ref{set2})
to zeroth order of $1/M_P^2$.  

Now we calculate the next correction in $\Phi$.  The  corrections after solving Eq. (\ref{ode}) to leading order  
in $1/M_P^2$ is obtained to be
\ba
\Phi_1^{(1)} &=&\frac{1}{30 \lambda M_P^2t^5} 
\Big[ 4 k^2t^2 \Big( k t \cos (kt) - \sin (kt) \Big) Ci(2kt)+ 4 k^2 t^2 \Big( \cos (kt) + kt  \sin (kt) \Big )Si(2kt) \nonumber   \\ 
&& ~~~~~~~~~~~~~~~~~~~~~~~ +3 k^2 t^2 \sin(kt)+  23 k t \cos (kt)-  23  \sin(kt)  \Big]
\ea
and
\ba
\Phi_2^{(1)}& =&\frac{1}{30 \lambda M_P^2t^5} 
\Big[ -4 k^2t^2 \Big( \cos (kt)+kt \sin (kt) \Big) Ci(2kt)+ 4 k^2 t^2 \Big( kt \cos (kt)- \sin (kt) \Big )Si(2kt) \nonumber   \\ 
&& ~~~~~~~~~~~~~~~~~~~~~~-3 k^2 t^2 \cos(kt)+  23 k t \sin (kt)+ 23  \cos(kt)  \Big]
\ea
in which $Si(x) \equiv \int_{0}^{x} dy \sin(y)/y$,  $Ci(x)\equiv \gamma+\ln(x)+\int_{0}^{x}dy (\cos y-1)/y$ and $\gamma$ is the Euler number. Having obtained $\Phi_i= \Phi_i^{(0)} +   \Phi_i^{(1)}$ we can also calculate 
$\calR_i$ using Eq. (\ref{R-Phi2}). However, it is more instructive to look at the super-horizon limit 
of these solutions,  $\lambda k  M_P^2 | t|^3 \ll 1$ . For the first mode we obtain
\ba
\Phi_1 \simeq  \left( \frac{-1}{3}+  \frac{1}{9 \lambda M_P^2 t^2}\right) k^3 
+ \left( \frac{t^2}{30} + \frac{112 - 60 ( \gamma + \ln ( 2 k t)  }{1350  \lambda M_P^2 }  \right) k^5
\quad \quad  (\sqrt {\lambda} k  M_P t^2 \ll 1) \, ,
\ea
and
\ba
\calR_1 \simeq \frac{k^3}{3} + \left( -\frac{t^2}{18} + \frac{-17 + 12 ( \gamma +  \ln ( 2 k t)) }{270 \lambda M_P^2}  \right) k^5 
  \qquad \qquad  (\sqrt {\lambda} k  M_P t^2 \ll 1) \, .
\ea
In particular note that $\Phi_1 \simeq (-1+ \frac{1}{3 \lambda M_P^2 t^2}) \calR_1$ as anticipated from
Eq. (\ref{set1}). As expected, this mode satisfies the results of  \cite{Weinberg:2008zzc, Weinberg:2003sw}.

Now, let us look at the second mode in the super-horizon limit obtaining
\ba
\Phi_2 \simeq  \left( \frac{1}{t^3}+  \frac{23}{30 \lambda M_P^2 t^5}\right)  
+ \left( \frac{1}{2 t} + \frac{17 - 8 ( \gamma + \ln ( 2 k t)  }{60  \lambda M_P^2 t^3 } 
\right) k^2
\qquad \qquad  ( \sqrt {\lambda} k  M_P t^2 \ll 1 )
\ea
and
\ba
\label{R2-conformal}
\calR_2 \simeq  -\left( \frac{1}{3t} + \frac{7}{90 \lambda M_P^2 t^3} \right) k^2   
~~~~~~~\qquad \qquad \qquad \qquad  (\sqrt {\lambda} k  M_P t^2 \ll 1) \, .
\ea
In the mathematical limit in which $k=0$, we obtain $\calR_2 =0$ and $\Phi_2 \propto \frac{H}{a}$ in exact agreement with the results of \cite{Weinberg:2008zzc, Weinberg:2003sw}. In the physical  super-horizon limit
in which $\sqrt {\lambda} k  M_P t^2 \ll 1$ while  $k$ is held fixed we observe the $1/t$ grows of $\calR_2$ in super-horizon limit. We see that the situation here is very similar to discussions in sub-section \ref{Phi-method}. 
We also comment that  the $1/t$ growth of $\calR$ on super-horizon scales was also observed  in the model of Galilean Genesis \cite{Creminelli:2010ba}. 

It is also instructive to understand how the proof \cite{Weinberg:2008zzc, Weinberg:2003sw} is violated in 
pseudo conformal universe in the method discussed in sub-section \ref{R-method}. As we noticed there, the 
key place to look for is the Poisson equation. Let us start with the original Poisson equation (\ref{Poisson}) yielding for pseudo conformal  model
\ba
\label{Poisson3}
\left( \frac{-1}{\lambda\,  t^4} + M_P^2 k^2 \right) \Phi = \frac{1}{2}( -\dot \phi \delta \dot \phi + \ddot \phi \delta \phi ) \, .
\ea
In the proof of \cite{Weinberg:2008zzc, Weinberg:2003sw}  the mathematical super-horizon limit
corresponds to $k=0$ independent of how large $M_P$ is. However, similar to argument mentioned after 
Eqs. (\ref{example}) and (\ref{Poisson2}), this limit is ambiguous here. This is because  
in this model  gravity is  assumed to be very weak so we work in the limit $M_P \rightarrow \infty$.  Therefore, in order to be safe,  we shall keep both terms in big bracket in Eq. (\ref{Poisson3}). The rest of analysis go exactly as in sub-section \ref{R-method} and we obtain the second order differential equation for $\calR$ given in Eq. (\ref{R-eq}). Note the interesting fact that  in Eq. (\ref{R-eq}) no
factor of $M_P$ appears so no ambiguity in taking $k\rightarrow 0$ while $M_P \rightarrow \infty$ arises now.
In addition  $a(t)$ is very slow-changing and the $\epsilon$-dependence is  the same for both terms in 
Eq. (\ref{R-eq}).  Therefore, the mathematical super-horizon limit $k \rightarrow 0$ is justified
in Eq. (\ref{R-eq}). In this limit, the two independent solutions are given as in Eq. (\ref{R-sol}) represented by
constants $C_1$ and $C_2$. The first mode is the constant mode as expected. Now for the second  mode we obtain
\ba
\calR_2 =  C_2 \int \frac{dt }{a^3 \epsilon}  \simeq \frac{-C_2}{9 \lambda M_P^2} \frac{1}{t} \, .
\ea
Interestingly, we see again that $\calR_2 \propto \frac{1}{t}$ as obtained in Eq. (\ref{R2-conformal}).

\bigskip


To summarize,  in this work we have revisited the celebrated Weinberg theorem in cosmological perturbation theory. The theorem states that there always exists two adiabatic scalar modes which are constant on super-horizon scales. Despite its wide applicability, however there are known examples in literature which violate this theorem. We have concentrated on loopholes in some technical assumptions  which are violated 
in models of non-attractor inflation, fluid inflation, solid inflation and pseudo conformal universe. 

We have seen  that the theorem in \cite{Weinberg:2008zzc, Weinberg:2003sw} can be violated in two different ways. The obvious way is when there is non-analytic relation in terms of the wave-number  $k$ 
in Einstein fields equations. This situation was already  anticipated in  \cite{Weinberg:2008zzc, Weinberg:2003sw}. The case of solid inflation is a specific example in which $\pi^S$ is non-analytically related to $\zeta$ via $\pi^S \propto \zeta/k^2$.  However, the more non-trivial examples are the cases in which
some parameters of the background, like the slow-roll parameter $\epsilon$, show strong time-dependence in which  the mathematical treatment of the super-horizon  limit $k\rightarrow 0$ is ambiguous as we discussed after 
Eqs. (\ref{example}) and (\ref{Poisson2}).   This is the case in non-attractor inflation, fluid inflation and 
in pseudo conformal model. In the first two examples  $\epsilon $ falls off like $1/a^6$ and the combination $k^2/a^2 \epsilon$ appearing in Poisson  equation diverges even on super-horizon scales. In the latter example  $\epsilon \sim M_P^2 t^2 \gg 1$ showing a strong time-dependence.

\vspace{0.7cm}

{\bf Acknowledgments:}  We would like to thank P. Creminelli, J. Khoury, M. Mirbabayi, M. H. Namjoo, 
M. Sasaki, G. Tasinato  and
M. Zaldarriaga  for useful discussions and correspondences.  We also thank ICTP for hospitality during `` First ICTP Advanced School on Cosmology"  where this work was in progress.

{}

\end{document}